\documentclass[twocolumn,showpacs,aps,prd,floatfix]{revtex4}

\newif\ifpdf\pdffalse

\usepackage{graphicx}
\usepackage{dcolumn}
\usepackage{bm}

\newcommand{\BABARPubYear}    {04}
\newcommand{\BABARPubNumber}  {06}

\newcommand{\SLACPubNumber} {10439}

\newcommand{\boldq} {\mbox{\boldmath$q$}}
\newcommand{\boldQ} {\mbox{\boldmath$Q$}}

\input babarsym

\begin{document}

\preprint{\babar-PUB-\BABARPubYear/\BABARPubNumber}
\preprint{SLAC-PUB-\SLACPubNumber}

\begin{flushleft}
\babar-PUB-\BABARPubYear/\BABARPubNumber\\
SLAC-PUB-\SLACPubNumber\\[10mm]
\end{flushleft}

\title{\large \bf A Measurement of the Total Width, the
 Electronic Width, and the Mass of the {\boldmath $\Upsilon(10580)$} Resonance}

%
\author{B.~Aubert}
\author{R.~Barate}
\author{D.~Boutigny}
\author{F.~Couderc}
\author{J.-M.~Gaillard}
\author{A.~Hicheur}
\author{Y.~Karyotakis}
\author{J.~P.~Lees}
\author{V.~Tisserand}
\author{A.~Zghiche}
\affiliation{Laboratoire de Physique des Particules, F-74941 Annecy-le-Vieux, France }
% ...
\collaboration{The \babar\ Collaboration}
\noaffiliation

\date{\today}

\begin{abstract}
We present a measurement of the parameters of the $\Upsilon(10580)$
resonance based on a dataset collected with the \babar\ detector at
the SLAC \pep2\ asymmetric $B$ factory. We measure the total width
$\Gamma_{\rm tot} = (20.7\pm1.6\pm2.5) \mev$, the electronic
partial width $\Gamma_{ee} = (0.321\pm0.017\pm0.029) \kev$ and the
mass $M = (10579.3\pm0.4\pm1.2) \mevcc$.
\end{abstract}


\maketitle

\section{Introduction}
\label{sec:Introduction}

The $\Upsilon(10580)$ resonance is the lowest mass $b\bar{b}$ vector state
above open-bottom threshold that decays into two $B$ mesons. 
The total decay width $\Gamma_{\rm tot}$ of the
$\Upsilon(10580)$ is therefore much larger than the widths of the
lower mass $\Upsilon$ states, thereby allowing a direct measurement
of $\Gamma_{\rm tot}$ at an $e^+e^-$ collider. 
Although the state has been known for almost 20 years, its mass and width
have been known only with relatively large uncertainties, and with
central values from different experiments showing substantial variation
\cite{ARGUS,CLEO1,CLEO2,CUSB}.  We present new measurements of the
mass, the total width, and the electronic widths of the $\Upsilon(10580)$ with
improved precision.

\section{Experiment and Data}

The data used in this analysis were collected with the \babar\
detector at the \pep2\ storage ring \cite{PEP2}.  The data set
comprises three energy scans of the $\Upsilon(10580)$ and one scan
of the $\Y3S$ resonance.  The \pep2\ $B$ factory is a high-luminosity
asymmetric $e^+e^-$ collider designed to operate at a center-of-mass
(CM) energy around $10.58\gev$. 

The PEP-II energy is calculated from the values of the currents of the power supplies for 
the magnets in the ring. Every major magnet in the ring has been measured in the 
laboratory and a current ($I$) vs.\ magnetic field ($B$) curve is determined for each magnet. 
The curve is a 4th order polynomial 
fit to the measured data. Many of the ring magnets are connected in series as strings with 
a single power supply. For the high-energy ring (HER) the bend magnets are in two strings of 96 magnets 
each. The $I$ vs.\ $B$ curve for a particular magnet string is then the average of the measured 
curves of the magnets in the string. The HER bend magnets are sorted according to field 
strength at a fixed $I$ so that we have the following layout: high-medium-low then low-medium-high 
\cite{peplay}.  The power supplies are controlled by zero-flux transductors with 
each supply having a primary and a secondary transductor. The transductor accuracy is 
on the order of $10^{-5}$ and the secondary transductor is used to check the primary 
transductor.

When an energy scan is being made the CM energy 
is changed by changing the energy of the 
high-energy beam, while the low-energy beam is left unchanged. 
The energy of the HER is adjusted by increasing the 
current in all of the large magnet power supplies (main dipoles and all quadrupoles but no 
skew quadrupoles) by a calibrated amount based on the $I$ vs.\ $B$ curves for the power 
supplies. The small orbit-correctors in the beam are not changed. The beam orbit is 
monitored to ensure the orbit is not changing during an energy scan. Other variables that 
affect the beam energy via the RF frequency are also held constant. In the first 
energy scan PEP-II experienced problems with one or more RF stations in the HER. 
These stations (of which there were five at the time) add discrete amounts of energy to the 
beam at the location of the RF station to compensate for the beam-energy loss due to 
synchrotron radiation emission around the ring. If one or more stations are off due to 
problems, the actual beam energy at the collision point can change by a small amount, which 
depends on the station that was turned off \cite{PepCMenergy}.
 
In order to minimize magnet hysteresis effects, the ring magnets are standardized by 
ramping the magnets to a maximum current setting, then to zero current four times. This 
was also done before the $I$ vs.\ $B$ curves were measured as a function of increasing 
magnetic field. The ring energy is lowered to the lowest energy point of the scan and then 
the magnets are standardized. Energy scans are always done in the direction of increasing 
magnetic field.

\babar\ is a solenoidal detector optimized for the asymmetric beam
configuration at \pep2. Charged-particle momenta are measured in a
tracking system consisting of a five-layer, double-sided silicon
vertex tracker (SVT) and a 40-layer drift chamber (DCH) filled with a
mixture of helium and isobutane, operating in a $1.5$-T
superconducting solenoidal magnet. The electromagnetic calorimeter
(EMC) consists of 6580 CsI(Tl) crystals arranged in a barrel and forward
endcap. A detector of internally
reflected Cherenkov light (DIRC)
provides separation of pions, kaons and protons.
Muons and long-lived neutral hadrons are
identified in the instrumented flux return (IFR), composed of
resistive plate chambers and layers of iron.
A detailed
description of the detector can be found in Ref.~\cite{BabarNim}. 

\section{Resonance Shape}

The $\Upsilon(10580)$ resonance parameters can be determined by
measuring the energy dependence of the cross section $\sigma_{b\bbar}$
of the reaction $\epem \rightarrow \Upsilon(10580)
\rightarrow \BB$ in an energy interval around the resonance mass. 
The cross section of this process, neglecting radiative corrections and
the beam-energy spread, is given by a relativistic Breit-Wigner
function
\begin{equation}
\sigma_0(s) = 12\pi \frac{\Gamma^0_{ee}\Gamma_{\rm tot}(s)}{(s-M^2)^2 + M^2\Gamma_{\rm tot}^2(s)},
\label{eq:purereso}
\end{equation}
where $\Gamma^0_{ee}$ is the partial decay width into \epem,
$\Gamma_{\rm tot}$ is the total decay width, $M$ is the mass of the
resonance, and $\sqrt{s}$ is the CM energy of the $\epem$
collision. The partial decay width $\Gamma^0_{ee}$ is taken as constant
and the approximation $\Gamma_{\rm tot}(s) \approx
\Gamma_{\Y4S\rightarrow B\Bbar}(s)$ is used.

\begin{figure}[t]
\ifpdf
  \includegraphics[width=0.99\columnwidth]{gammaqpc.pdf}
\else
  \includegraphics[width=0.99\columnwidth]{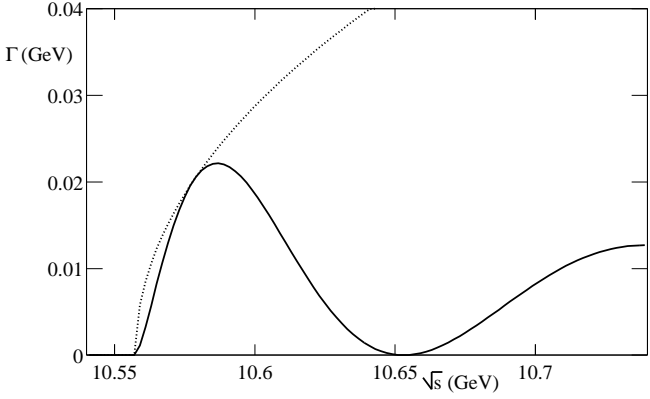}
\fi
\caption{The decay width $\Gamma_{\Y4S\rightarrow B\Bbar}(s)$ for the QPC model
(solid line) compared to phase space alone (dotted line).
Due to the proximity to the threshold, the width rises
steeply.  However, the overlap integral of the 4S Upsilon state with the
1S $B$-meson states vanishes three times due to the nodes of the 4S wave function, and pushes
$\Gamma(s)$ down.}
\label{fig:gams}
\end{figure}

Since the $\Upsilon(10580)$ is so close to the threshold for $B\Bbar$ production,
its width $\Gamma_{\rm tot}(s)$ is expected to vary strongly with energy $\sqrt s$.
It rises from zero at $\sqrt s = 2 m_B$, but its behavior beyond that depends 
on decay dynamics.
The quark-pair-creation model (QPCM) \cite{QPCM} is used to describe
these dynamics. It is a straightforward model where the $b$
and $\bar{b}$ quarks from the bound state, together with a
quark-antiquark pair created from the vacuum, combine to form a
$\Bbar$ and a $B$ meson. The matrix element for this decay is given by
a spin-dependent amplitude and an overlap integral of the
$\Upsilon(10580)$, treated as a pure $4S$ state.
\begin{equation}
\Gamma_{\Y4S\rightarrow B\Bbar}(s) = {1\over 8\pi} \left\vert
g_{BB\Upsilon} \sum_{m=\pm1} I_4(m,\boldq)
\right\vert^2 {q(s)\over s}
\label{eq:width}
\end{equation}
where $m$ is the 3-component of the
$\Upsilon$ spin. 
The overlap integral of the $\Y{n}S$ state with two
$B$ mesons
\begin{eqnarray}
I_n(m,\boldq) &= \int &Y_1^m(2 \boldq - \boldQ)
\,\psi_{\Y{n}S}(\boldQ) 
\,\psi_B(\boldQ - h\boldq)\times \nonumber\\
&&{}\times \psi_{\Bbar}(-\boldQ + h\boldq)
\,\mathrm{d}^3 Q
\end{eqnarray}
where $\boldq$ is the momentum vector of the $B$ meson,
and
$h = 2m_b/(m_b + m_q)$ \cite{Yaouanc}.
The calculation based on the harmonic oscillator wave function
$$
\psi(\boldq) = \left(R^2 \over \pi\right)^{3\over 4} e^{-R^2 q^2/8}
$$
for the 1S state
yields 
\begin{eqnarray}
I_4(m,\boldq) &=\displaystyle{ \sqrt{1\over 35} }\biggl[&
14 R^2 {\partial \over \partial R^2}
+ 16 R^4 \left(\partial \over \partial R^2\right)^2 \nonumber\\
&&{}+ {16\over3} R^6 \left(\partial \over \partial R^2\right)^3
\biggr]  I_1(m,\boldq)
\label{eq:ovlapp}
\end{eqnarray}
with $R = R_{\Y4S}$ and
\begin{eqnarray}
I_1(m,\boldq) &=& {8 \sqrt{6} \over \pi^2} 
\left(R R_B^2 \over R^2 + 2 R_B^2\right)^{3/2}
\left(1 - {h R_B^2 \over R^2 + 2 R_B^2}\right) \times\nonumber\\
&&{}\times \exp\left(-{R^2 R_B^2 h^2 q^2 \over 4(R^2 + 2 R_B^2)}\right) 
\mbox{\boldmath$\epsilon$}(m)\cdot \boldq
\end{eqnarray}
We use the approximation with harmonic-oscillator wave
functions provided by the ARGUS
collaboration \cite{ARGUS}, i.e., 
the Hamiltonian
\begin{eqnarray}
{\cal H}& = &m_b + m_q - {(m_b + m_q) \nabla^2 \over 2 m_b m_q}\nonumber\\
&&{}+ 0.186\gev^2 r - {4\alpha_s\over 3\,r} - 0.802\gev
\end{eqnarray}
with
$\alpha_s = 0.35(0.42)$ for the $\Y4S$ ($B$),
$m_b = 5.17\gev$ and $m_q = 0.33\gev$, 
where they obtain as a minimum of $\langle \psi|{\cal H}|\psi\rangle$
the values
$R = R_{\Y4S} = 1.707\gev^{-1}$,
$R_B = 2.478\gev^{-1}$.
The resulting $\Gamma_{\Y4S\rightarrow B\Bbar}(s)$ is shown in Figure~\ref{fig:gams}
and compared to the behaviour of spin-0 pointlike particles.
The fact that the $\Upsilon$- and $B$-mesons are extended objects modifies
the shape significantly.

The uncertainty of this model is parametrized 
as one constant $g_{BB\Upsilon}$, representing
the coupling of the $\Y4S$ to a $B\Bbar$ pair,
and is absorbed in the fit to the data 
by the free total width 
$\Gamma_{\rm tot} = \Gamma(M^2)$, assuming $\Gamma_{\rm tot} \approx \Gamma_{B\Bbar}$.
The free parameters of this model are hence the mass $M$ and the
width $\Gamma_{\rm tot}$.

The resonance shape is significantly modified by QED corrections
\cite{QEDcor1,QEDcor2}. The cross section including radiative
corrections of {\cal O}($\alpha^3$) is given by
\begin{equation}
\tilde{\sigma}(s) = \int\limits^{1-4m_e^2/s}_0 \sigma_0(s-s\kappa) \beta \kappa^{\beta-1}(1+\delta_{\rm vert} +\delta_{\rm vac}) \, \mathrm{d}\kappa,
\label{correctedform}
\end{equation}
where $\kappa = \frac{2E_{\gamma}}{\sqrt{s}}$ is the scaled energy of the
radiated photon, $\beta =
\frac{2\alpha}{\pi}(\ln \frac{s}{m_e^2}-1)$, and $\delta_{\rm vert} =
\frac{2\alpha}{\pi}(\frac{3}{4} \ln \frac{s}{m_e^2} -1
+\frac{\pi^2}{6})$ is the vertex correction.  The vacuum polarization
of the photon propagator $\delta_{\rm vac}$ is absorbed in the
physical partial width $\Gamma_{ee} \approx \Gamma^0_{ee}(1+\delta_{\rm vac})$ \cite{Gamee}.

\begin{figure}[t]
\ifpdf
  \includegraphics[width=0.99\columnwidth]{sigmaevol.pdf}
\else
  \includegraphics[width=0.99\columnwidth]{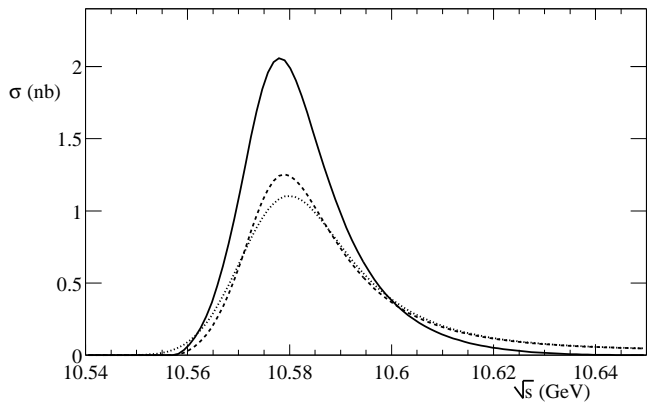}
\fi
\caption{Cross section without (solid line) and including (dashed line)
initial photon radiation.  Further broadening from the beam energy spread
leads to the shape given by the dotted line.}
\label{fig:smear}
\end{figure}

A second modification of the cross section arises from the beam-energy
spread of \pep2. Averaging over the $\epem$ CM energies $\sqrt{s'}$,
which are assumed to have a Gaussian distribution around the mean value
$\sqrt{s}$ with a standard deviation $\Delta$, results in a cross
section of:
\begin{equation}
\sigma_{b\bbar}(s) = \int \tilde{\sigma}(s') \frac{1}{\sqrt{2\pi} \Delta} 
\exp \left( -\frac{(\sqrt{s'}-\sqrt{s})^2}{2\Delta^2}\right) \,\mathrm{d}\sqrt{s'}.
\label{finalformula}
\end{equation}
Extraction of $\Gamma_{\rm tot}$ from the observed resonance shape
requires knowledge of the energy spread $\Delta$.  The spread is
measured from a scan of the narrow \Y3S resonance.
Both effects are illustrated in Figure~\ref{fig:smear}.

\section{Data Analysis}

The strategy of this analysis is to determine the shape of the
$\Upsilon(10580)$ resonance from three energy scans in which the cross
section is measured from small data samples at several CM energies
near the resonance.  These are combined with a precise measurement of
the peak cross section from a high-statistics data set with a well
understood detector efficiency taken close to the peak in the course
of $B$-meson data accumulation.  

\subsection{Event Selection}

The visible hadronic cross section measured from the number of
hadronic events $N_{\rm had}$ and the luminosity $L$ is related to
$\sigma_{b\bbar}$ via
\begin{equation}
\sigma^{\rm vis}(s) \equiv \frac{N_{\rm had}}{L} = \varepsilon_{b\bbar} \sigma_{b\bbar}(s) + \frac{P}{s},
\label{visform}
\end{equation}
where $\varepsilon_{b\bbar}$ is the detection efficiency for
$\Upsilon(10580) \to B\Bbar$. The parameter $P$ describes the
amount of background from non-$B\Bbar$ events, which are dominantly $\epem
\rightarrow q\bar{q}$.

Any selection of hadronic events will have backgrounds from two
classes of sources. Processes such as $e^+ e^- \rightarrow q \qbar
(\gamma)$, $e^+ e^- \rightarrow e^+ e^-
e^+ e^-$ or $e^+ e^- \rightarrow
\tau^+ \tau^- (\gamma)$ all
have cross sections $\sigma \propto 1/s$ with corrections that are
negligible over the limited energy range of each scan. This permits
describing this class of backgrounds in a fit to the data by one
parameter $P$. The second class of backgrounds originates from
two-photon processes $\gamma \gamma \rightarrow {}$hadrons or beam-gas
interactions, which do not scale in a simple way with energy. The
latter process even depends on the vacuum in the beam pipe rather than
on the beam energy.  This kind of background cannot be taken into
account in the fit of the resonance. Therefore the event selection
must reduce this background to a negligible level.

Hadronic events are selected by exploiting the fact that they have a
higher charged-track multiplicity $N_{\rm ch}$ and have an event-shape
that is more spherical than background events.  Charged tracks are
required to originate from the beam-crossing region and the event
shape is measured with the normalized second Fox-Wolfram moment $R_2$
\cite{foxwolfram}. Additional selection criteria are applied to reduce the beam-gas and
$\gamma\gamma$ backgrounds. The particular criteria for the analysis
of the $\Y3S$ scan data, the peak cross section measurement, and the
$\Upsilon(10580)$ scan are described in the paragraphs below.

\subsection{Luminosity Determination}

The luminosity is measured from $e^+e^- \rightarrow \mu^+ \mu^-$
events. These events are required to have at least one pair of charged
tracks with an invariant mass greater than $7.5\gevcc$. The
acolinearity angle between these tracks in the CM has to be smaller than
10 degrees to reject cosmic rays. At least one of the tracks must have
associated energy deposited in the calorimeter. Bhabha events are
vetoed by requiring that none of the tracks has an associated energy
deposited in the calorimeter of more than 1 GeV.

\subsection{Calibration Using the $\Y3S$ Resonance}

\begin{figure}[t]
\ifpdf
  \includegraphics[height=5.5cm]{y3sscan.pdf}
\else
  \includegraphics[height=5.5cm]{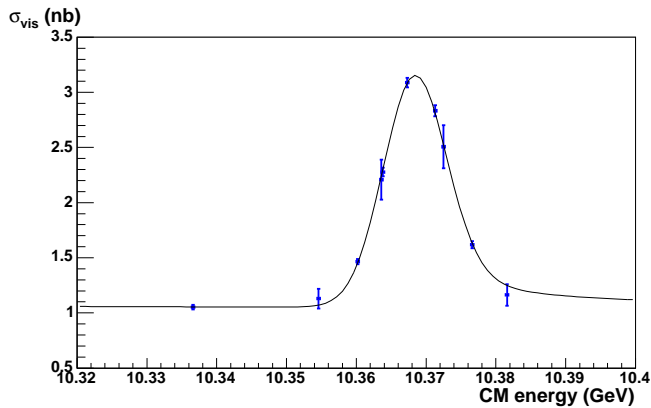}
\fi
\caption{Visible cross section after event selection vs.\ the uncorrected CM energy for the \Y3S 
resonance scan. The line is the result of a fit.}
\label{fig:y3s}
\end{figure}

The \Y3S scan taken in November 2002 consists of ten cross section
measurements performed at different CM energies. The energies are
obtained from the settings of the \pep2 storage ring. The visible
cross section $\sigma_{\rm vis}$ is measured for each energy. The \Y3S
decays have higher multiplicity and are more isotropic than the
continuum background, which allows us to select
\Y3S events with requirements similar to those used for the $B\Bbar$ selection.  
In particular, the criteria $R_2 < 0.4$ and $N_{\rm ch} \ge
3$ are used to select hadronic events. Additionally, the invariant mass
of all tracks combined is required to be greater than 2.2 \gevcc.

The branching fraction of the \Y3S into $\mu^+\mu^-$ 
corresponds to a cross section of $\sim 0.1
\nb$ for resonant muon-pair production. Therefore, the luminosity is
determined from Bhabha events for the data points of the \Y3S scan.
Figure \ref{fig:y3s} shows the data points and the result of a fit.

The Breit-Wigner function (\ref{eq:purereso}) of the \Y3S resonance is
approximated by a delta function because the width of the \Y3S,
$\Gamma^{\rm 3S}_{\rm tot} = (26.3\pm3.4)\kev$ \cite{PDG2002}, is very
small compared to the energy spread of \pep2. The cross section is
related to the visible cross section via equation (\ref{visform}),
which is fitted to the data points.  The free parameters of the fit
are the \Y3S mass $M_{3S}^{fit}$, the energy spread $\Delta$, the
parameter $P$ describing the background, and $\varepsilon
{\Gamma_{ee}\Gamma_{\rm had}\over\Gamma_{\rm tot}}$, where
$\varepsilon$ is the efficiency for selecting \Y3S decays.  The result
of the fit including the statistical errors are $$
\begin{array}{rcl}
\Delta &=& (4.44\pm0.09)\mev,\\
M_{\rm 3S}^{\rm fit} &=& (10367.98 \pm 0.09)\mevcc,\\
\end{array}
$$ 
with $\chi^2/{\rm dof} = 2.2/6$.  Sources of a systematic
uncertainty in the fit results are potential variations of the
detector and trigger performance during the $\Y3S$ scan and the
precision ($\pm 0.20\mev$) of the determination of the energy
differences between the scan points. In total, the systematic
uncertainty is estimated to be $0.17 \mev$ and $0.15 \mevcc$ for the
energy spread and $\Y3S$ mass, respectively.

The observed shift of 0.12\% between the fitted \Y3S mass $M_{\rm
3S}^{\rm fit}$ and the world average of $(10355.2\pm0.5)$ \mevcc
\cite{vepp4} is used to correct the \pep2\ CM energies. The machine
energy spread is extrapolated to $10580.0 \mevcc$ by scaling the
spread of the high-energy beam with the square of its energy,
resulting in $\Delta = (4.63\pm0.20)\mev$.  An extrapolation of the
spread of the low-energy ring is not necessary, because its energy was
held constant.  The energy spread during two of the three
$\Upsilon(10580)$ scans was $0.2\mev$ larger.
This larger spread was caused by a wiggler that
ran at full power till late February 2000. Since this date it runs at only
10\% of its full power, which reduces its influence on the spread.

\subsection{The $\Upsilon(10580)$ Peak Cross Section}

The $b\bar{b}$ cross section at the peak of the $\Upsilon(10580)$
resonance is determined from the energy dependence of
$\sigma_{b\bar{b}}$ measured from a high-statistics data set.
These data were taken between October 1999 and June 2002
close to the peak, at energies between $10579$ and
$10582\mev$.  They comprise an integrated luminosity
of $76\fb^{-1}$, much larger than the typical $0.01\fb^{-1}$ of a scan.
The
cross section $\sigma_{b\bbar}$ is given by
\begin{equation}
  \sigma_{b\bbar} = \frac{N_{\rm had} - N_{\rm \mu\mu} \cdot R_{\rm off} \cdot r}{\varepsilon_{b\bbar}'L},
\end{equation}
where $N_{\rm \mu\mu}$ is the number of muon pairs, $R_{\rm off}$ is
the ratio of hadronic events to muon pairs below the resonance,
$\varepsilon_{b\bbar}'$ is the efficiency for selecting $\BB$ events,
and $r$ is a factor close to unity, estimated from Monte Carlo
simulation, that corrects for variations of cross sections and
efficiencies with the CM energy.

We apply cuts on track multiplicity, $N_{\rm ch} \ge 3$, and on the
event-shape, $R_2 < 0.5$, to select these hadronic events. Events from
$\gamma\gamma$ interactions and beam-gas background are reduced by
selecting only events with a total energy greater than $4.5
\gev$. Beam-gas interactions are additionally reduced by requiring that
the primary vertex of these events lies in the beam collision region.

The peak cross section is determined from this long run on resonance.
To take into account the tiny variations of the hadronic cross section
close to the maximum, we
fit a third-order polynomial to the cross sections
$\sigma(e^+ e^- \to B\Bbar)$ as a function of uncorrected energy
(the energy of the peak position is not used in this analysis, 
instead the $\Upsilon(10580)$
mass
is determined solely from the short-time scans as descibed below).
This results in a peak value
of $(1.101\pm0.005\pm0.022) \nb$. 
The second error is systematic and
includes as dominant contributions uncertainties in the efficiency
$\varepsilon_{b\bbar}'$, calculated from Monte Carlo simulation,
and in the luminosity determination.

\subsection{The Three $\Upsilon(10580)$ Scans}

\begin{table}[t]
\caption{Data points of the 1st scan of the $\Upsilon(10580)$ resonance. The cross sections are not efficiency corrected. The energies of this scan are shifted by a constant offset relative to the energy scale of the other two scans. The offset is a free parameter in the simultaneous fit to all three scans. The CM energy spread during this scan was $\Delta = 4.83\mev$.}
\begin{ruledtabular}
\begin{center}
\begin{tabular}{cccc}
 & CM energy ($\mev$) & $\sigma_{vis}$ (nb) & \\
\hline 
 & 10518.2 & $0.777 \pm 0.060$ & \\
 & 10530.0 & $0.868 \pm 0.048$ & \\
 & 10541.8 & $0.828 \pm 0.046$ & \\
 & 10553.7 & $0.762 \pm 0.050$ & \\
 & 10565.5 & $0.933 \pm 0.044$ & \\
 & 10571.4 & $1.203 \pm 0.037$ & \\
 & 10577.3 & $1.4466 \pm 0.0207$ & \\
 & 10583.3 & $1.706 \pm 0.064$ & \\
 & 10589.2 & $1.615 \pm 0.122$ & \\
 & 10595.3 & $1.291 \pm 0.117$ & \\
 & 10601.3 & $1.091 \pm 0.101$ & \\
\end{tabular}
\end{center}
\label{tab:y4sdata99}
\end{ruledtabular}
\end{table}

\begin{table}[ht]
\caption{Data points of the 2nd scan of the $\Upsilon(10580)$
resonance. The cross sections are not efficiency corrected. The CM
energy spread during this scan was $\Delta = 4.83\mev$. The energy
correction obtained from the $\Y3S$ scan is applied to the CM
energies.}
\begin{ruledtabular}
\begin{center}
\begin{tabular}{cccc}
 & corrected CM energy ($\mev$) & $\sigma_{vis}$ (nb) & \\
\hline 
 & 10539.3 & $0.9429 \pm 0.0282$ & \\
 & 10571.6 & $1.452 \pm 0.054$ & \\
 & 10576.7 & $1.756 \pm 0.050$ & \\
 & 10579.6 & $1.730 \pm 0.044$ & \\
 & 10584.7 & $1.650 \pm 0.063$ & \\
 & 10591.4 & $1.457 \pm 0.043$ & \\
 & 10604.3 & $1.0686 \pm 0.0295$ & \\
\end{tabular}
\end{center}
\label{tab:y4sdata00}
\end{ruledtabular}
\end{table}

\begin{table}[ht]
\caption{Data points of the 3rd scan of the $\Upsilon(10580)$
resonance. The cross sections are not efficiency corrected. The CM
energy spread during this scan was $\Delta = 4.63\mev$. The energy
correction obtained from the $\Y3S$ scan is applied to the CM
energies.}
\begin{ruledtabular}
\begin{center}
\begin{tabular}{cccc}
 & corrected CM energy ($\mev$) & $\sigma_{vis}$ (nb) &  \\
\hline 
 & 10539.6 & $0.9775 \pm 0.0249$ & \\
 & 10570.4 & $1.5236 \pm 0.0293$ & \\
 & 10579.4 & $1.857 \pm 0.040$ & \\
 & 10579.4 & $1.850 \pm 0.033$ & \\
 & 10589.4 & $1.656 \pm 0.038$ & \\
\end{tabular}
\end{center}
\label{tab:y4sdata01}
\end{ruledtabular}
\end{table}

\begin{table*}[ht]
\caption{Comparison of the results obtained from a fit to the three
$\Upsilon(10580)$ scans using a non-relativistic Breit-Wigner function
with an energy independent total decay width (1st row) and the
quark-pair-creation model (2nd row) to describe the resonance shape,
respectively. The quark-pair-creation model describes the energy
dependence of the total decay width close to the open bottom threshold
taking spatial features of the $\Upsilon(4S)$ meson wave function into
account. We therefore use this model for our measurement, while the
fit with a non-relativistic Breit-Wigner function is used as an
estimate for the model uncertainties.}
\begin{ruledtabular}
\begin{center}
\begin{tabular}{lccccc}
 & $\Gamma_{\rm tot}$ [\mev] & $\Gamma_{ee}$ [\kev] & $B_{ee} \times
 10^5$ & M [\gevcc]& $\chi^2/{\rm dof}$ \\
\hline 
 non-rel. Breit-Wigner, $\Gamma_{\rm tot} = {}$const&  $17.9\pm1.3$ & $0.288\pm0.015$ & $1.61\pm 0.04$ & $10.5796\pm0.0004$ & 15.4/14 \\
 quark-pair-creation model & $20.7\pm1.6$ & $0.321\pm0.017$ & $1.55\pm0.04$ & $10.5793\pm0.0004$ & $18.3/14$ \\
\end{tabular}
\end{center}
\label{tab:fitresults}
\end{ruledtabular}
\end{table*}

\begin{table*}[ht]
\caption{Summary of systematic uncertainties}
\begin{center}
\begin{tabular}{l|crc|clc|c|crc}
\hline
\hline
 & \multicolumn{3}{c|}{$\delta\Gamma_{\rm tot}$ (MeV)} & \multicolumn{3}{c|}{$\delta \Gamma_{ee}$ (keV)} & $\delta B_{ee}\times10^{5}$ & \multicolumn{3}{c}{$\delta M$ (\mevcc)} \\
\hline
 model uncertainty & & 1.4 & & & 0.017 & & 0.03 & & 0.1 & \\
 systematic bias by single data point & & 2.0 & & & 0.022 & & 0.04 & & 0.3 & \\
 uncertainty of energy spread & & 0.5 & & & 0.0024 & & 0.03 & & $<$  0.1  \\
 uncertainty of peak cross section & & $<$ 0.1 & & & 0.006 & & 0.03 & & $<$ 0.1 & \\
 long term drift of energy scale & \multicolumn{3}{c|}{-} &  \multicolumn{3}{c|}{-} & - & & 1.0 & \\
 error on $M_{\rm \Y3S}$ &  \multicolumn{3}{c|}{-}  &  \multicolumn{3}{c|}{-}  & - & & 0.5  & \\ 
\hline
 total error & & 2.5 && & 0.029 & & 0.07 & & 1.2  \\
\hline
\hline
\end{tabular}
\end{center}
\label{tab:sys}
\end{table*}

\begin{table}[ht]
\caption{Central values of the $\Upsilon(10580)$ resonance parameters including their statistical errors and correlation coefficients of the fit to the three $\Upsilon(10580)$ scans. Any combination of two of the 
 three parameters $\Gamma_{\rm tot}$, $\Gamma_{ee}$ and $B_{ee}$ can be used as free parameters in
 the fit.}
\begin{ruledtabular}
\begin{center}
\begin{tabular}{c|cc|ccc}
 & value obtained from fit & & $\Gamma_{ee}$ & $B_{ee}$ & $M$  \\
\hline
$\Gamma_{\rm tot}$ & $(20.7\pm1.6)\mev$ & & 0.996 & -0.980 & 0.206 \\
$\Gamma_{ee}$ & $(0.321\pm0.017)\kev$ & & & -0.961 & 0.186 \\
$B_{ee} $ & $(1.55\pm0.04)\cdot 10^{-5}$ & & & & -0.226 \\
$M$ & $(10579.3\pm0.4)\mev$ & & & & \\
\end{tabular}
\end{center}
\label{tab:corelmatrix}
\end{ruledtabular}
\end{table}

The $\Upsilon(10580)$ scan consists of three scans around the
resonance mass taken in June 1999, January 2000 and
February 2001. Hadronic events are selected by requiring $N_{\rm ch}
\ge 4$ and $R_2 < 0.3$. The background from beam-gas and
$\gamma\gamma$ interactions is reduced by the cut $E_{\rm tot}-|P_z| >
0.2 \sqrt{s} $, where $E_{\rm tot}$ is the total CM energy calculated
from all charged tracks and $P_z$ is the component of the total
CM momentum of all charged tracks along the beam axis.

\begin{figure}[t]
\begin{center}
\ifpdf
\includegraphics[height=11.5cm]{y4sscans5.pdf}
\else
\includegraphics[height=11.5cm]{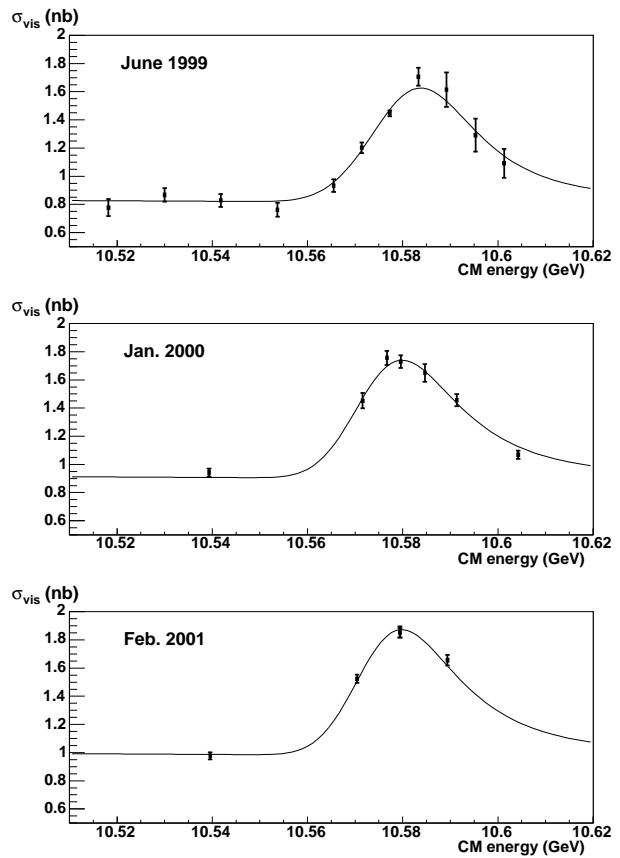}
\fi
\caption{Visible cross section after event selection vs.\ CM energy for the three $\Upsilon(10580)$ 
scans. The lines are the result of a simultaneous fit to all three scans.}
\label{fig:y4sfit}
\end{center}
\end{figure}

The data points ($\sigma^{\rm vis}_i$,$\sqrt{s_i}$)
are listed in Tables~\ref{tab:y4sdata99}--\ref{tab:y4sdata01}.  They are 
shown in
Fig.~\ref{fig:y4sfit} together with a fit based on Eq.
(\ref{visform}).  The CM energies of the $\Upsilon(10580)$ scans from
Jan.\ 2000 and Feb.\ 2001 are corrected using the shift obtained from
the \Y3S fit. This is not possible for the CM energies of the scan
from June 1999. 
In this scan, which took several days, it was possible to have the energy drift while data were
being collected at a scan point. 
These drifts have been monitored and the average energies are
corrected to $\pm0.05\mev$, so that point-to-point energy
variations are still negligible.
The absolute scale, however, can not precisely be calibrated
to that of the $\Y3S$ scan.
For this reason a mass shift between that scan and the
later two scans has to be included as a free parameter into the
fit. The other free parameters are the total width $\Gamma_{\rm tot} =
\Gamma_{\rm tot}(M^2)$, the electronic width $\Gamma_{ee}$, the mass
$M$ of the $\Upsilon(10580)$ and for each scan the background
parameter $P$ and the efficiency $\varepsilon_{b\bar{b}}$. The
efficiencies can be free parameters in the fit since we fix 
the peak cross section for each
scan to the value obtained from the on-resonance data set. The energy
spread of the collider is fixed to $4.63\mev$ for the scan of February
2001 and to $4.83\mev$ for the other two scans.  
Note that the branching fraction
$B_{ee} = {\Gamma_{ee} / \Gamma_{\rm tot}}$ is not an independent parameter.
The fit results 
for the resonance parameters
are given in Table~\ref{tab:corelmatrix} together with the correlation matrix.
The other fit parameters agree with expectations. 

\subsection{Systematic Uncertainties}

We treat the $\Upsilon(10580)$ resonance as a $4S$ state, but its
shape is slightly modified by mixing with the $\Upsilon_1(3D)$ and
possibly other states as well as by coupled-channel effects at higher
energies above the thresholds for $BB^*$ and $B^* B^*$ production
\cite{ono}.  An analysis of the energy region around the
$\Upsilon(10580)$ that includes all possible states and decay channels
is not possible because of the limited energy range of \pep2\ and the
lack of more detailed theoretical models.  Instead, we treat the
$\Upsilon(10580)$ as a resonance well enough isolated from other peaks
to be described in a model using a pure $4S$ state. This is one reason
to omit data taken at CM energies well above the $BB^*$ threshold.
Another reason is the fact that details of the meson wave functions
become more significant at higher energies, as can be learned from
Figure~\ref{fig:gams}.

To estimate the effect of our model we use the
width of the resonance shape defined by the full width at half maximum
(FWHM) as an alternative definition for $\Gamma_{\rm tot}$. The FWHM
is obtained replacing (\ref{eq:purereso}) with a non-relativistic 
Breit-Wigner function
with constant width $\Gamma_{\rm tot} = {}$const in the fit 
to the data points.  This would be the approach
when nothing is known about the nature of the resonance.
The results are summarized in Table~\ref{tab:fitresults}. 
The difference in the 
fit results tells us the effect of our more refined description.
We assume a model uncertainty of 50\%, 
i.e., we take half of 
the difference for each fit parameter as an estimate of the
model uncertainties.

A systematic bias in the fit results could be caused by detector
instabilities or an incorrect energy measurement during a scan. This effect
is estimated by excluding single data points from the fit. The maximum
shift for each fit parameter is taken as a systematic error.
 
The \Y3S scan and the $\Upsilon(10580)$ scans were spread over a
period of three years.  A systematic error of $1.0\mev$ is assigned to
the mass measurement due to drifts in the beam energy determination
between the $\Upsilon(10580)$ scans and the \Y3S scan that are not
reflected in the beam energy corrections. These drifts are caused by
changes of the beam orbit and ring circumference. Another systematic
error on the mass measurement arises from the uncertainty in the mass
of the \Y3S. The systematic error caused by the uncertainty of the
energy spread of the collider is estimated by varying the energy
spread used in the fit procedure for all three $\Upsilon(10580)$ scans
by its uncertainty of $\pm0.20\mev$. Long-term fluctuations of the
energy spread are taken into account by varying the energy spread of
single scans in the fit by $\pm 0.1\mev$. The quadratic sum of both
contributions is listed in Table~\ref{tab:sys}.  In addition the
systematic error due to the uncertainty in the peak cross section is
included. The systematic uncertainties due to energy dependences of
the event selection efficiencies are found to be negligible.

\section{Summary}

Our final results are
$$
\begin{array}{rcl}
\Gamma_{\rm tot} &=& (20.7 \pm 1.6 \pm 2.5)\mev, \\
\Gamma_{ee} &=& (0.321\pm0.017\pm0.029)\kev, \\
B_{ee} &=&(1.55\pm0.04\pm0.07)\cdot 10^{-5}, \\
M &=& (10579.3\pm0.4\pm1.2)\mevcc. \\
\end{array}
$$
The measurements of the total width and mass are improvements in
precision over the current world averages \cite{PDG2002}. 

\section{Acknowledgments}
\label{sec:Acknowledgments}

We appreciate helpful discussions with Alain Le Yaouanc.
We are grateful for the excellent luminosity and machine conditions
provided by our \pep2\ colleagues, 
and for the substantial dedicated effort from
the computing organizations that support \babar.
The collaborating institutions wish to thank 
SLAC for its support and kind hospitality. 
This work is supported by
DOE
and NSF (USA),
NSERC (Canada),
IHEP (China),
CEA and
CNRS-IN2P3
(France),
BMBF and DFG
(Germany),
INFN (Italy),
FOM (The Netherlands),
NFR (Norway),
MIST (Russia), and
PPARC (United Kingdom). 
Individuals have received support from the 
A.~P.~Sloan Foundation, 
Research Corporation,
and Alexander von Humboldt Foundation.


\begin{thebibliography}{99}

\bibitem{ARGUS} ARGUS Collaboration, H. Albrecht {\it et al.}, Z. Phys. {\bf C65} 619 (1995). 

\bibitem{CLEO1} CLEO Collaboration, D.~Besson {\it et al.}, Phys.\ Rev.\ Lett.\  {\bf 54}, 381 (1985).
 
\bibitem{CLEO2} CLEO Collaboration, C.~Bebek {\it et al.}, Phys.\ Rev.\ {\bf D36}, 1289 (1987).
 
\bibitem{CUSB}  CUSB Collaboration, D.~M.~Lovelock {\it et al.}, Phys.\ Rev.\ Lett.\  {\bf 54}, 377 (1985).  

\bibitem{PEP2} PEP-II: An Asymmetric B Factory. Conceptual Design Report, SLAC-R-418, LBL-PUB-5379 (1993).

\bibitem{peplay}U. Wienands {\it et al.},  
IEEE Proc. of the 1995 Particle Accelerator Conference
and International Conference on High-energy Accelerators, Dallas, Texas, May 1995,
Piscataway, NJ, IEEE, vol. 1, p. 530--532 (1996). 

\bibitem{PepCMenergy} M. Sullivan, M. Donald, and M. Placidi, 
IEEE Proc. of the 2001 Particle Accelerator Conference, Chicago, Illinois, June 2001,
eds. P. Lucas, S. Webber,
Piscataway, N.J., IEEE, 2001, vol. 5, p. 3570--3572 (2001).

\bibitem{BabarNim} \babar\ Collaboration, B. Aubert {\it et al.}, Nucl.
  Instr. and Methods {\bf A479}, 1 (2002).
  
\bibitem{QPCM} A. Le Yaouanc, L. Oliver, O. Pene, J.-C. Raynal, Phys. Rev. {\bf D8} 2223 (1973);
 A. Le Yaouanc, L. Oliver, O. Pene, J.-C. Raynal, Phys. Lett. {\bf B71} 397 (1977); 
 S. Ono, Phys. Rev. {\bf D23} 1118 (1981).

\bibitem{Yaouanc} A. Le Yaouanc, priv.\ communication (1999); $h$ differs by a factor 2 from \cite{ARGUS}.

\bibitem{QEDcor1} E. A. Kuraev, V. S. Fadin, Sov. J. Nucl. Phys. {\bf 41}, 466 (1985).

\bibitem{QEDcor2} J. P. Alexander, G. Bonvicini, P. S. Drell, R. Frey, Phys. Rev. {\bf D37}, 56 (1988).

\bibitem{Gamee} D. Albert, J. Marciano, D. Wyler, Z. Parsa, Nucl. Phys. {\bf B166} 460 (1980).

\bibitem{foxwolfram}G. C. Fox, S. Wolfram,
 Phys.\ Rev.\ Lett.\  {\bf 41}, 1581 (1978) and Nucl. Phys. {\bf B149}, 413 (1979).

\bibitem{PDG2002} Particle Data Group, S. Eidelman {\it et al.}, Phys. Lett. {\bf B592}, 1 (2004).

\bibitem{vepp4} A. S. Artamonov {\it et al.}, Phys. Lett. {\bf B474} 427 (2000).

\bibitem{ono} K. Heikil\"a, N. A. T\"ornqvist, S. Ono, Phys.\ Rev.\ {\bf D29}, 110 (1984).

\end{thebibliography}
\end{document}